\documentclass[twocolumn]{revtex4}
\usepackage{amssymb,epsf}
\usepackage{latexsym}
\usepackage{xcolor}
\usepackage{epsfig}
\usepackage{float}

\begin{document}

\title{On the thermodynamics of reconciling quantum and gravity}
\author{H. Moradpour$^1$\footnote{h.moradpour@riaam.ac.ir}, S. Jalalzadeh$^2$\footnote{shahram.jalalzadeh@ufpe.br}, Umesh Kumar Sharma$^3$\footnote{sharma.umesh@gla.ac.in}}
\address{$^1$ Research Institute for Astronomy and Astrophysics of Maragha (RIAAM), University of Maragheh, P.O. Box 55136-553, Maragheh, Iran\\
$^2$ Departamento de F\'{i}sica, Universidade Federal de
Pernambuco, Recife-PE, 52171-900, Brazil\\
$^3$ Department of Mathematics, Institute of Applied Sciences and
Humanities, GLA University Mathura 281406, Uttar Pradesh, India}

\begin{abstract}
Is thermodynamics consistent with the quantum gravity
reconciliation hypothesis [A. G. Cohen et al. Phys. Rev. Lett. 82,
4971 (1999)], which establishes holographic dark energy models?
Here, we have attempted to address this issue in the affirmative
by concentrating on the first law of thermodynamics.
\end{abstract}


\maketitle
\section{Introduction}

One of the most challenging issues in the standard model of
cosmology is the origin of dark energy (DE). Of all the hypotheses
given to represent the universe's evolution, the standard method,
which includes the ordinary Friedmann equations produced by
solving the Einstein field equations with a constant termed the
cosmological constant (CC), is the most useful. The choice of CC
is due to its simplicity and general relativity's (GR)
compatibility with other areas of physics, such as particle
physics, thermodynamics, and observations. Despite significant
advancements in modern physics, the nature of CC remains unsolved.
Another contentious subject is the connection between quantum
physics and gravity. It may be solved after we discover the
ultimate theory of gravity, or it may help us find out the
promised theory of gravity.


In order to solve this paradigm and reconcile quantum mechanics
and gravity, Cohen et al. \cite{HDE} proposed
\begin{eqnarray}\label{1}
\rho_\Lambda<\frac{S}{2L^4},
\end{eqnarray}
where $\rho_\Lambda$ denotes the vacuum energy density
\cite{HDE,HDE01,HDE1,HDE2,HDE3}. Indeed, this bound is also
obtained by conditioning that the vacuum energy
($E_\Lambda\approx\rho_\Lambda L^3$), stored in volume $L^3$,
should not exceed the energy of the same size black hole
($E\approx L$) i.e. \cite{HDE,shey}
\begin{eqnarray}\label{0}
E_\Lambda<E.
\end{eqnarray}
 In the framework of GR, where we have $S=\frac{A}{4}\sim
L^2$ (the Bekenstein entropy-area relation) and $E=L/2$ for the
Misner--Sharp energy of a black hole, one can easily see that both
inequalities (\ref{1}) and~(\ref{0}) lead to
$\rho_\Lambda<L^{-2}$.

On the other hand, in dynamic situations, it appears that the
apparent horizon is a proper casual boundary of the universe
\cite{cons,cons1,cons2,cons3,Cai2,CaiKimt}. Motivated by these
statements and supposing that a black hole of the same size as the
cosmos also satisfies the Bekenstein entropy, it is proposed that
the vacuum energy stored in the universe, bounded by the apparent
horizon, may form a model for DE.
Here is where the Holographic Dark Energy (HDE) first manifested
itself, based on introducing $\rho_\Lambda=C^2\frac{S}{L^4}$ as
DE,
\cite{HDE01,HDE1,HDE2,HDE3}. 
It should be emphasized that Cohen's approach only assumes that
the universe would collapse into itself if its radius equals its
Schwarzchild radius; it does not state that the entropy of the
black hole and the universe are equivalent because their contents
and structures differ.

Due to the shortcomings of primary HDE (PHDE) \cite{HDE3}, various
modifications have subsequently been introduced and classified
into three classes, including $i$) corrections to the Bekenstein
entropy due to various effects like quantum considerations, $ii$)
other boundaries like particle horizon, $iii$) interactions
between various sectors of cosmic fluid, $iv$) the nature of the
proportionality coefficient $C^2$ (constant or variable)
\cite{RevH}, and $v$) alternative options for the Bekenstein
entropy obtained by using the generalized statistics such as
Tsallis HDE (THDE) \cite{THDE}, Sharma--Mittal HDE (SMHDE)
\cite{SMHDE}, Kaniadakis and new Tsallis HDE models (KHDE and
NTHDE, respectively) \cite{KHDE,P:2022amn,Pandey:2021fvr}.

Parallel to the deep connection between thermodynamics and gravity
\cite{J1,T1,J11}, it is also shown that the Friedmann equation can
be rewritten as the first law of thermodynamics. Although in the
dynamic situation two temperatures are proposed for the horizon,
called the Hayward--Kodama temperature \cite{cons,cons1,cons2} and
the Cai--Kim temperature \cite{cons3,CaiKimt}, both of them
satisfy the first law of thermodynamics on the apparent horizon
and respect the Friedmann equation
\cite{cons,cons1,cons2,cons3,CaiKimt,AHEPD,ijtp}. Consequently, on
the one hand, the Bekenstein entropy bound is respected by the
quantum field theory \cite{sred}, and on the other hand, the
revealed deep connection between thermodynamics and gravity
\cite{J1,T1,J11} confirms a preliminary expectation, i.e., the
true physical statements and laws should confirm each other.


As a result, because vacuum energy is an essential element of the
cosmos, the question of whether Cohen et al.'s idea actually meets
the principles of thermodynamics emerges. Indeed, the physical
statements must concur, which means that a suggestion like that of
Cohen et al. must be consistent with thermodynamic laws. The
investigation of the existence and degree of this compatibility is
the foundation of this study.


Our primary purpose here is to establish a link between
thermodynamics and the idea of Ref. \cite{HDE}. We will first
describe related earlier studies in the next section to accomplish
this goal. The third section will investigate the compatibility of
bound (\ref{1}) with the first law of thermodynamics. The final
part contains a summary.

\section{Thermodynamics inspires holographic bounds}

In the framework of thermodynamics, one may justify Eq.~(\ref{1})
as:
\begin{itemize}
  \item Total energy of quantum fields in a vacuum is estimated as $E_\Lambda\sim\rho_\Lambda
  V$, in which $V$ denotes the cosmos volume \cite{sh1,sh2}.
  \item This energy should be smaller than the energy of the same size black
  hole ($E$), as it has not turned into a black hole.
  \item Bearing the Euler relation, ($E=TS$), \cite{CALLEN} in mind and accepting $T\sim\frac{1}{L}$
  as the temperature of the same size black hole, the energy of the black hole with entropy $S$
  may be estimated as $E=TS\sim\frac{S}{L}$.
  \item Finally, since $V\sim L^3$, by combining the above steps,
  one can easily reach $\rho_\Lambda<\frac{S}{L^4}$. Consequently, in this manner, one does not
  need to assume $S\sim L^2$~(\ref{1}), and the result is proposed for all
  functions of $S(L)$.
\end{itemize}

Regarding the above justification, it is concluded that the RHS of
Eq.~(\ref{1}) is based on the Euler relation, which is not a law:
The Euler relation is a crucial statement in classical
thermodynamics that expresses internal energy as the sum of the
products of canonical pairs of extended and intense variables. The
name ``Euler relation'' refers to the fact that this relationship
is obtained using Euler's homogeneous function theorem.
Subsequently and motivated by attempts like~\cite{J1,T1,J11}, one
may arrange to use the Clausius relation, $dE=TdS$, which is a
form of the first law of thermodynamics and should be valid on
casual boundaries \cite{CALLEN}. Therefore, as a step forward to
relate holographic bound to thermodynamics, it may be argued that
a proper holographic bound should satisfy the $dE=TdS$ relation
instead of the Euler equation. This assertion motivates us to
consider the vacuum energy stored in differential volume, $dV$.
Relying on the proposal of Cohen et al., its energy should be
smaller than the same size black hole. This leads us to consider
\cite{RHDE}
\begin{eqnarray}\label{6}
\rho_\Lambda<\frac{dE}{dV}\Rightarrow
\rho_\Lambda<T\frac{dS}{dV}\sim\frac{\frac{dS}{dA}}{L^2},
\end{eqnarray}
where $dE=TdS$ is assumed. Moreover, the thermodynamic
justification of Eq.~(\ref{1}) as well as $T\sim\frac{1}{L}$
(temperature of the same size black hole) \cite{HDE} has also been
adopted. It means that Eq.~(\ref{6}) may be considered as a
thermodynamic inspiration of Eq.(\ref{1}). It is also easy to see
that the use of Eq.~(\ref{6}) generates the outcome of
Eq.~(\ref{1}) for $S=S_B$.


Using~(\ref{6}) and the R\'{e}nyi entropy, also motivated by the
probable quantum aspects of gravity \cite{epl}, R\'{e}nyi HDE
(RHDE) has recently been introduced in \cite{RHDE}. In this model,
there are two entropies: 1) the FRW universe's boundary entropy
(horizon entropy) and 2) the entropy of a black hole (the
same-size black hole) used to scale the vacuum energy. Apart from
the limitations of this argument \cite{manoh}, this is an attempt
to propose a thermodynamic approach to reconcile quantum and
gravity, resulting in an HDE model with appropriate behavior (see
Ref.~\cite{RHDE} and its citations).

Because of the anonymous nature of DE, article~\cite{ijtp} claims
that we should not rule out the probability of entropy creation
via DE. Thus, while the Friedmann equation is correct, the total
entropy may diverge from Bekenstein entropy due to the presence of
DE and its interaction with the environment. In this manner,
starting from the first Friedmann equation \cite{ijtp}
\begin{equation}\label{2}
H^2+\frac{\kappa}{a^2}=\frac{8\pi}{3}\rho,
\end{equation}
accepting the energy definition $dE=\rho dV$, (where
$\rho=\rho_1+\rho_D$ in which $\rho_D$ and $\rho_1$ denote the
energy densities of DE and the remaining parts of cosmic fluid,
respectively), and using the first law, one can find
\begin{eqnarray}\label{3}
S_A=S_B+\frac{1}{6}\int A^2 d\rho_D,
\end{eqnarray}
 whenever there is not any mutual interaction between DE
and other parts of cosmic fluid \cite{ijtp}. Here, $S_B$
represents Bekenstein entropy ($\frac{A}{4}$), and $S_A$ is the
total entropy. If the DE nature is identical to $\rho_1$, then one
reaches $S_A=S_B$ \cite{ijtp}. If one uses Tsallis entropy
\cite{tsallis}, or fractional--fractal entropy
\cite{Jalalzadeh:2022uhl,Jalalzadeh:2021gtq}, and Eq.~(\ref{3}),
then the obtained DE seems to be able to justify DE \cite{ijtp}.
Indeed, Eq.~(\ref{3}) says that while the Universe dynamics is
governed by the Friedmann equations, DE, and its interaction with
other parts of the cosmos may produce entropy. In this manner, one
can find $\rho_D$ ($S_A$) by finding $S_A$ ($\rho_D$). In the
presence of interaction ($Q$), one obtains \cite{AHEPD}
\begin{equation}\label{4}
S_A=S_B+\frac{1}{6}\int A^2 \big[d\rho_D+Qdt\big].
\end{equation}
 More related studies can also be found in
\cite{cana,mitra,mitra1,mitra2,mitra3}. As a side note, if one
does not decompose $\rho$ to $\rho_1$ and $\rho_D$, then the
ordinary solution \cite{AHEPD,ijtp}
\begin{eqnarray}\label{5}
\rho=-6\int \frac{S_A^{\prime}}{A^2}dA,
\end{eqnarray}
is obtained by employing the first law of thermodynamics on the
cosmos boundary surface, and using the energy-momentum
conservation law. Here, $S_A$ denotes the boundary entropy,
$S_A^{\prime}\equiv\frac{dS_A}{dA}$, and one can easily get
Eq.~(\ref{1}) by hiring $S_B$ in calculating Eq.~(\ref{5}).

For a non-interacting cosmic fluids, using Eq.~(\ref{3}), one
finds
\begin{eqnarray}\label{8}
\rho_D=6\int \frac{dS_A}{A^2}-\frac{3}{2A}.
\end{eqnarray}
Now, as an example, let us consider \cite{power}
\begin{eqnarray}\label{9}
S_A=S_B+\alpha A^\mu,
\end{eqnarray}
where $\alpha$ and $\mu$ are unknown coefficients. Utilizing
Eq.~(\ref{8}), we obtain
\begin{eqnarray}\label{10}
\rho_D=BL^{2\mu-4},
\end{eqnarray}
 where $B=(4\pi)^{\mu-2}\frac{6\alpha\mu}{\mu-2}$ and
$A=4\pi L^2$, in which $L$ denotes the radius of the boundary, is
used to obtain this result. Mathematically, it is THDE obtained by
employing Eq.~(\ref{1}), and Tsallis  \cite{THDE} (or
fractional--fractal \cite{Jalalzadeh:2021gtq,Jalalzadeh:2022uhl})
entropy.


In Ref.~\cite{manoh}, authors urge modifying PHDE (the
proposal~(\ref{1})) as
\begin{eqnarray}\label{7}
\rho_\Lambda\propto 6\int\frac{dS}{A^2}-\frac{3}{2A},
\end{eqnarray}
which is obviously compatible with Eq.~(\ref{8}). While $S_A$
represents the total entropy of the cosmic boundary ($S_B$ plus
the entropy amount produced by DE) \cite{ijtp}, $S$ (in
Eq.~(\ref{7})) is the entropy of the same size black hole (the
same size as the cosmos). Therefore, one may deduce that if the
entropy of the same-size black hole is equal to the sum of the
Bekenstein entropy and the entropy produced by DE, then
Eq.~(\ref{7}) should reduce to Eq.~(\ref{8}). Consequently, if
vacuum energy is involved in DE, then a preliminary cautious
estimation can be $\rho_\Lambda\propto\rho_D$ (take into account
that $i$) $S_A\neq S$, $ii$) this probability that all of the
vacuum energy, stored in the cosmos, may not have a role in the
current accelerated phase of the universe, and $iii$) DE is not
entirely made up of vacuum energy). Accepting
$\rho_\Lambda\propto\rho_D$, one can start from Eq.~(\ref{8}) to
reach Eq.~(\ref{7}). Hence, we may accept Eq.~(\ref{8}) as the
thermodynamic motivation of bound~(\ref{7}).


\section{The First law of thermodynamics}

Prior to moving on, it's essential to keep in mind a few things.
\begin{itemize}
\item We always have a constant of integration at the last step
due to integration in Eqs.~(\ref{8}) and~(\ref{7}) (and comparable
relations). We usually disregard the integration constant
\cite{manoh}, although its existence occasionally enhances the
model \cite{prd}. Issues involving such constants are well-known,
such as the CC problem.
 \item The bound~(\ref{6}) is in a differential form which means it does not include the constant of integration, which is an advantage over (\ref{8}).
    \item It is assumed in the proposed thermodynamic explanation of Eq.~(\ref{1}) that $T\sim1/L$. It contradicts the standard effective field theory \cite{HDE} as well as gravitational theories in which the temperature of a black hole is proportional to its surface gravity \cite{pois}.
  \item Whereas Eq.~(\ref{1}) thermodynamic's reasoning is based on the Euler equation, the Clausius relation provides the foundation of Eq.~(\ref{6}). Nevertheless, the first law's involvement in Eq.~(\ref{1}) is unclear because the Clausius relation contains only some information encoded in the first law \cite{CALLEN}. The gravitational field equations satisfy this fundamental law, and any physical formulation is assumed to be compatible with it.
  \item The pressure of the same-size black hole is ignored in all of the alternatives considered. Moreover, the GR field equations on a black hole's horizon take the form of the first statement ($dE=TdS-PdV$ is fulfilled) \cite{J1,T1,J11}, implying that the contribution of the pressure of a black hole of the same size to its energy content should also be included.
\end{itemize}

The first law ($dE=TdS-PdV$) is more basic than the Euler equation
($E=TS-PV\equiv R$), as previously stated. To address the
difficulties raised, we suggest using the $dE=TdS-PdV$ version of
the first law to compute the energy of the same size black hole,
which is an extension of Eq.~(\ref{6}), which does not contain the
$PdV$ component. The question here is what happens to the
additional terms in the differential of the Euler equation when
compared to the first law. Mathematically, we have
\begin{eqnarray}\label{11}
dR=\underbrace{TdS-PdV}_{\textmd{the\ first\ law}=dE}+\ \ SdT-VdP,
\end{eqnarray}
\noindent while the Gibbs-Duhem equation ($S=V\frac{dP}{dT}$)
conditions $SdT-VdP=0$ leading to $dR=dE$ \cite{CALLEN}, meaning
there is no need to worry about the extra terms \cite{CALLEN}. The
vacuum energy stored in the volume $dV$ is equal to
$dE_\Lambda=\rho_\Lambda dV$, and as it has not turned into a
black hole, we should have $dE_\Lambda<dE=dR$, where $dE$ is the
energy of the same size black hole. It is nothing but the
proposal~(\ref{0}) in volume $dV$ and covers the original bound
$E_\Lambda<R$~(\ref{0}) since $dR=dE$. Consequently, we have
\begin{eqnarray}\label{12}
&&\frac{dE_\Lambda}{dV}<\frac{dE}{dV}\Rightarrow\rho_\Lambda<\frac{dE}{dV}=
T\frac{dS}{dV}-P,
\end{eqnarray}

It is worth noting that this finding is not limited to the
Schwarzschild characteristics of energy and entropy. Moreover, it
states that the entropy, pressure, and temperature of a black hole
of the same size should be estimated thermodynamically, and hence
the findings may change under different gravitational theories.

\subsection*{GR and Bekenstein entropy}

Consider the static spherically symmetric spacetime, whose horizon
located at radius $r=L$. The line element is given by
\begin{equation}\label{met2}
ds^{2}=-f(r)dt^{2}+\frac{dr^{2}}{f(r)}+r^{2}d\Omega^{2}.
\end{equation}
In the framework of GR, where horizon entropy obeys the Bekenstein
entropy and we have $P=\frac{1}{8\pi
r^2}[rf^\prime(r)-1]\big|_{r=L}$ for the black hole pressure on
the horizon \cite{pois}, Eq.~(\ref{12}) leads to
\begin{eqnarray}\label{13}
\rho_\Lambda<\frac{1}{2A}\Rightarrow\rho_\Lambda<L^{-2},
\end{eqnarray}
as the thermodynamic bound on the vacuum energy. Here,
$T=\frac{f^\prime(r)}{4\pi}$ (the Hawking temperature of horizon)
has been used. In the GR framework, both Eqs.~(\ref{1})
and~(\ref{13}) claim that $\rho_\Lambda$ is scaled as $L^{-2}$.
Therefore, the proposal of Cohen et al. is compatible with the
first law.

%
%
%
%
%
\section{Conclusion}
Physical theories and logic must be compatible with one another.
Consequently, thermodynamic predictions, for instance, should
match those of large quantum systems. In the context of the
quantum gravity reconciliation hypothesis, it is noteworthy to
discuss the $S\propto A^n$ relation for the horizon entropy
derived by working in the context of Tsallis statistics
\cite{tsallis} or addressing quantum gravity concerns
\cite{Jalalzadeh:2021gtq,Jalalzadeh:2022uhl,Rashki:2014noa,Jalalzadeh:2014jea,barrow}.
This suggests that two distinct driving forces produce the same
predictions, implying that non-extensivity and quantum aspects of
gravity might be connected \cite{epl,homa}.
Motivated by $i$) the above expectation, $ii$) the the deep
connection between thermodynamics and gravity, and $iii$) some
recent attempts to propose HDE models by relying on the proposal
of Cohen et al. and using the $dE=TdS$ form of the first law
\cite{ijtp,RHDE,manoh}, it is tried to find out the relationship
between thermodynamic laws and holographic bound introduced in
\cite{HDE}. In particular, we concluded that the proposal of Cohen
et al. is consistent with the first law if the pressure
contribution is taken into account.


\section*{Acknowledgments}
S.J. acknowledges financial support from the National Council for
Scientific and Technological Development -- CNPq, Grant no.
308131/2022-3.

\section*{Declaration of competing interest}
The authors declare that they have no known competing financial
interests or personal relationships that could have appeared to
influence the work reported in this paper.


\end{document}